# On the gallium experiment BEST-2 with a $^{65}$Zn source to search for neutrino oscillations on a short baseline

V.N. Gavrin, V.V. Gorbachev[*], T.V. Ibragimova, V.N. Kornoukhov

*Institute for Nuclear Research of the Russian Academy of Sciences*
*60[th] October Anniversary pr., 7a  Moscow 117312 Russia*

A.A. Dzhanelidze, S.B. Zlokazov, N.A. Kotelnikov

*Joint stock company "Institute of Nuclear Materials"*
*P.O. Box 29, Zarechny, Sverdlovsk region, , 624250 Russia*

A.L. Izhutov, S.V. Mainskov, V.V. Pimenov

*ISC «SSC RIAR»*
*Zapadnoe shosse, 9, Dimitrovgrad, Ulyanovsk region, 433510, Russia*

V.P. Borisenko, K.B. Kiselev, M.P. Tsevelev

*Federal State Unitary Enterprice "Mayak Production Association"*
*Lenin pr., 31, Ozersk, Chelyabinsk region, 456784 Russia*

[*]*e-mail: vvgor_gfb1@mail.ru*

**Abstract** – In the paper is considered the use of a $^{65}$Zn source in the BEST-2 gallium experiment to constrain the regions of the allowed oscillation parameters. The required activity of the $^{65}$Zn source for the BEST-2 experiment, its size, effect on the results of oscillatory measurements, as well as the possibility of production such a source are calculated. Schemes of measurements execution are considered.

## 1. Introduction

Currently several experiments are being prepared with various sources of neutrinos, to test the hypothesis of the existence of a sterile neutrino or the fourth neutrino mass eigenstate [1-4].

In this paper the possibilities of the BEST-2 experiment with an intense artificial neutrino source $^{65}$Zn on a two-zone gallium target [5, 6] are shown. The experiment BEST-2 is considered as a continuation and complementation of the BEST experiment with the 3 MCi $^{51}$Cr source [1,2].

## 2. The BEST experiment

The BEST experiment is preparing on the basis of the gallium-germanium neutrino telescope at the Baksan neutrino Observatory of the INR RAS (GGNT), which has been used since 1990 for solar neutrino measurements in the SAGE experiment [7].

In the GGNT neutrinos are registered through neutrino capture reaction $\nu_e + {}^{71}Ga \rightarrow {}^{71}Ge + e^-$, and the number of interactions is determined by the measured number of produced $^{71}$Ge atoms.

In the BEST experiment the search for oscillations on a short baseline will be made by analyzing the capture rates of electron neutrinos from an intense $^{51}$Cr source with an activity of 3 MCi on the nuclei of the gallium target [1]. The target containing 50 tons of liquid metal gallium is divided into two zones - the inner spherical and the outer cylindrical with a common center. The source of about 1 liter is placed in the center of both target zones through a tube passing from top along the axis of the cylindrical target zone. With the identical average neutrino pass length from the source in both zones of the target, i.e. an equal the gallium thickness in 4π-geometry, the capture rate of neutrinos in the absence of oscillations in both zones will be the same and equal to 65 day$^{-1}$ (the capture rate of solar neutrinos in the entire target is about 1 per day).

Oscillations change the neutrinos flavor and reduce the capture rate of electronic neutrinos. For oscillations on a short baseline the electronic neutrino survival probability is

$$P_{ee} = 1 - \sin^2 2\theta \cdot \sin^2\left(\frac{1.27 \cdot \Delta m^2 \cdot L}{E}\right), \quad (1)$$

where the neutrino energy E is measured in MeV and the path length L - in meters. Parameters – the amplitude of the oscillation $\sin^2 2\theta$ and the squared mass difference of the mass eigenstates $\Delta m^2$ (eV$^2$) – are characteristics of oscillations which must be found in the experiment.

The results of the previous experiments with short neutrino path lengths (accelerator, gallium with intensive artificial sources, reactor with distances up to 100 m) [8-13] indicate that the parameter $\Delta m^2$ has a value of the order of 1 eV$^2$. For neutrinos with energy of 1 MeV the oscillation length at $\Delta m^2 = 1$ eV$^2$ is 2.5 m, so in the BEST experiment at the thickness of each target zone of 60 cm can be expected that the neutrinos capture rates in the two zones will significantly differ due to oscillations. In the measurements, the capture rates are summarized over all distances within one zone:

$$R = \frac{\int P_{ee}(L) \cdot s(L) dL}{\int s(L) dL} \quad (2).$$

Here R is the ratio of the measured capture rate to the expected in the given zone of target. Function *s(L)* determines the relative amount of neutrino interactions at *L,* which is the distance between the point of emission neutrino in the source and the point of neutrino capture in the target. The effect of oscillations on the capture rate is averaged over the distances within each target zone.

The BEST experiment sensitivity to oscillations is determined by experimental errors. In the absence of oscillations the expected statistical error of measurements is 3.7% for each zone of the target and 2.6 % for the entire target.

A systematic error also is assumed of 2.6% for each zone and for the entire target. That leading to a total uncertainty, statistical plus systematic of 4.5% for each zone and of 3.7 % for the entire target. Additionally, the uncertainty of the neutrino capture cross section by $^{71}$Ga nuclei of +3.6 / -3.0 % should be included in the error [14].

At the same time, the possibility of determination the values of the parameter $\Delta m^2$ in the BEST experiment is restricted by ranges of $\Delta m^2$ values, at which the capture rates in the outer and inner zones of the target differ significantly (see Fig.7) [15]. To expand the range of $\Delta m^2$ values that can be determined in the measurements, the experiment BEST-2 is proposed under the scheme and on the basis of the BEST experiment with another neutrino source. Instead of the $^{51}$Cr a $^{65}$Zn source will be used [5,6]. The isotope $^{65}$Zn emits neutrinos with energy of 1.35 MeV, 1.8 times larger of the neutrinos energy from $^{51}$Cr (0.75 MeV). Below we consider the possibilities of the BEST-2 experiment, its expected results and compare with the results of the experiment with the $^{51}$Cr source.

## 3. $^{65}$Zn source

The decay scheme of $^{65}$Zn is shown in Fig.1.

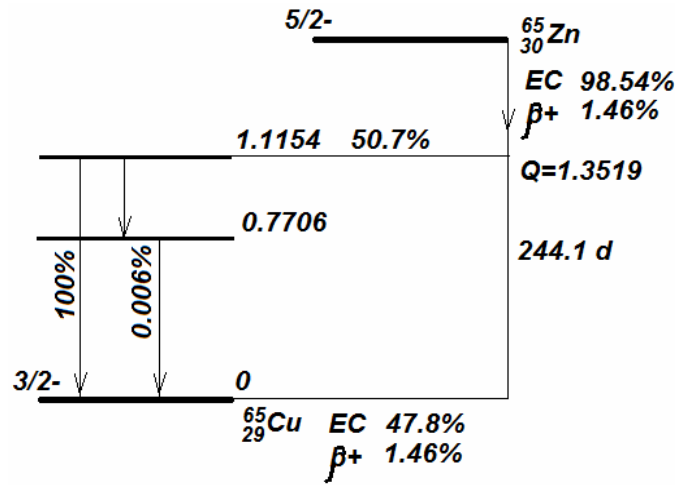

Fig.1. The decay scheme of $^{65}$Zn

Neutrinos with 1.35 MeV are emitted almost in half of the decays. Rest of half decays with the emission of 235 keV neutrinos, i.e. with the energy close to the capture threshold at gallium (233 keV), the cross-section of such captures is low.

The cross section for 1.35 MeV neutrino capture on $^{71}$Ga nuclei is about 3 times higher than for neutrinos with energy of 0.75 MeV from $^{51}$Cr [14]. Therefore, the expected neutrino capture rate in one target zone from the 3 MCi $^{65}$Zn source with the same size as the Chromium source, and also with the previous target zones configurations, will be equal to $n_0=108$ d$^{-1}$. Taking into account that the $^{65}$Zn lifetime is longer compared to $^{51}$Cr ($T_{1/2}$ = 244.1 and 27.7 days, respectively), measurements with $^{65}$Zn can be made for a longer time and the source activity can be noticeably lower to ensure of comparable statistics. Let's estimate what activity can be a $^{65}$Zn source.

## 4. Activity of the $^{65}$Zn source

The irradiation and extraction procedures from GGNT are well studied and adjusted in a long time in the experiment SAGE [7]. During the $t_1$ time the gallium target is irradiated (exposure) by a neutrino flux and the $^{71}$Ge atoms are produced. After $t_1$, during the $t_2$ time the produced atoms are extracted from the target for subsequent procedures of their counting. The irradiation time $t_1$ depends on the lifetime of $^{71}$Ge ($T_{1/2}$=11.43 days). For example, for solar neutrino extractions was taken exposure schedule (irradiations to solar neutrinos) with $t_1 = 30$ days and $t_2$ time was about 1 day.

If all the irradiations in the experiment are carried out within the same $t_1$ time period with identical $t_2$ dead time period between irradiations, then the collected statistics in the experiment is described by the expression:

$$N(m) = \frac{n_0}{\lambda_1 - \lambda_0} \cdot (e^{-\lambda_0 t_1} - e^{-\lambda_1 t_1}) \cdot \frac{1 - e^{-mB}}{1 - e^{-B}} \quad (3),$$

where $m$ is the number of exposures; $\lambda_1$ and $\lambda_0$ are the decay constants of the source and $^{71}$Ge respectively; $B = \lambda_1 \cdot (t_1 + t_2)$.

Fig.2 shows the dependences of the $N(m)$ total number of the collected $^{71}$Ge atoms for the sources $^{51}$Cr and $^{65}$Zn on the number of exposures for integer values $t_1$, at which $N$ is the maximal value for a given $m$. Fig.3 shows the values of $t_1$, which provide the maximal value $N$ for given $m$.

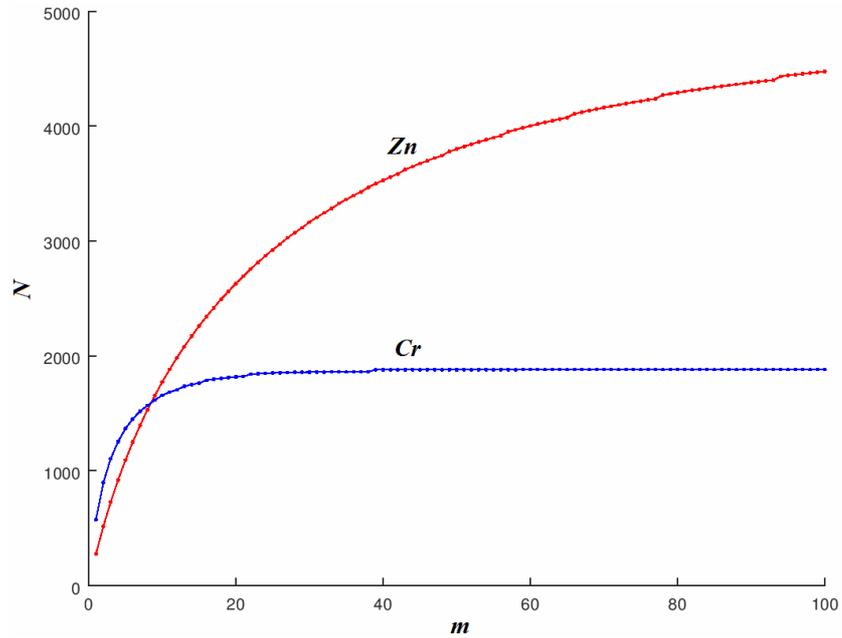

Fig.2. Dependence the number of events $N$ on the number of exposures $m$ for experiments with the $^{51}$Cr and $^{65}$Zn sources. The exposure time $t_1$ is various for different $m$, to provide the maximum number of events $N$.

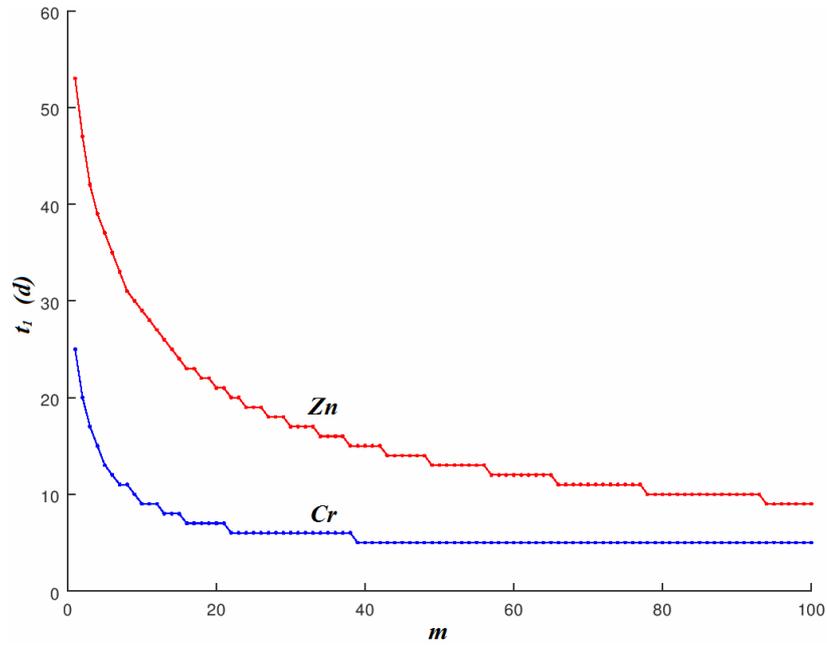

Fig.3. Dependence exposure time $t_1$ on the number of exposures $m$ for experiments with the $^{51}$Cr and $^{65}$Zn sources. $t_1$ take only integer values and vary with $m$ to obtain the maximum number of $N$ events.

Procedures of the BEST-2 experiment with the $^{65}$Zn source similar to ones of the BEST experiment with $^{51}$Cr source, and errors also will be comparable. The systematics of the experiments will be almost the same; the only difference is the background uncertainty of solar neutrinos that due to different times of one exposure $t_1$, and the number of exposures $m$ and due to different methods of measuring the sources activity. Background of solar neutrinos in the experiment BEST leads to an error, which is about 0.18% for the irradiation time $t_1$=9 days. For the increased by about 3 times irradiation time, t1 ~ 30 days, with the same number of irradiations $m$=10 an error will increase by ~1.8 times (the accumulation of $^{71}$Ge atoms in the target is proportional to $(1-e^{-\lambda t_1})$), i.e. to 0.3 %.

The statistics in the BEST experiment are determined by the total number of extracted $^{71}$Ge. For irradiation schedule with $m$=10 and $t_1$=9 days, the expected number of extracted $^{71}$Ge atoms in the absence of oscillations will be 1657. Let's determining a desired activity of the $^{65}$Zn source, to obtain the same number of events in BEST-2.

For the $^{65}$Zn source, which has a longer lifetime, it is convenient to use the measurement schedule with $t_1 \approx 30$ days, which corresponds to the solar neutrino measurement schedule of SAGE. Fig. 4 shows the dependences of the total number of the extracted $^{71}$Ge atoms on the number of the target exposures $N(m)$ with a fixed time of one exposure $t_1$=30 days for various source activities.

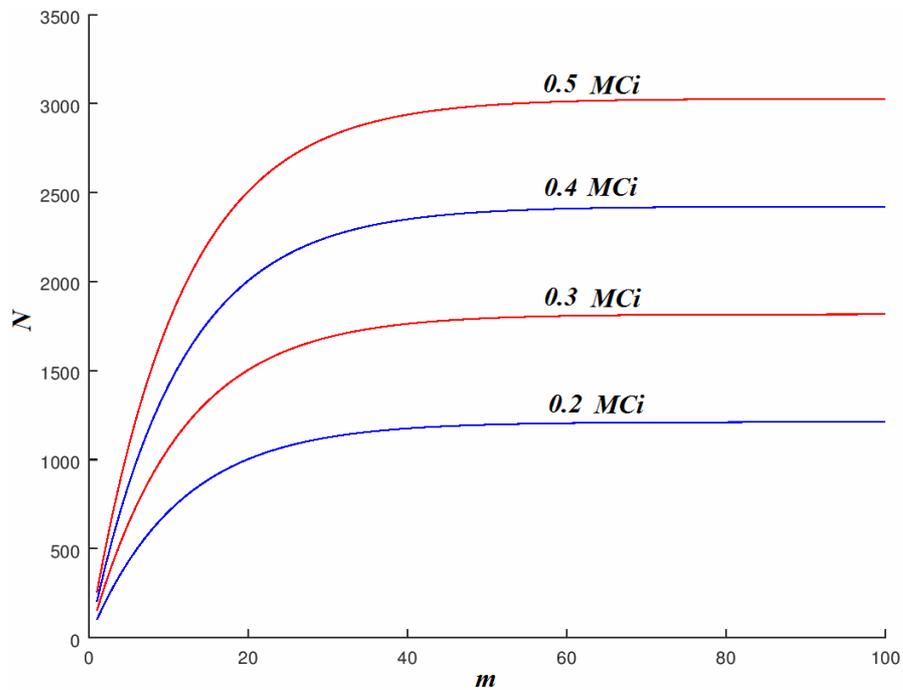

Fig.4. The accumulation of statistics $N(m)$ in experiments with $^{65}$Zn source with an activity of 0.2 to 0.5 MCi. The abscissa axis is the number of exposures $m$ with $t_1$=30 days.

The total amount of events N = 1657 can be reached already with a $^{65}$Zn source of 0.33 MCi, i.e. almost 10 times lower the $^{51}$Cr source activity in the BEST experiment. The number of exposures in this case will be $m \sim 28$, and the experiment will continue $t \approx (t_1+t_2)\cdot m \approx 868$ days = 2.38 years.

## 5. Source production

The $^{65}$Zn source can be produced by irradiating of enriched up to ~94% $^{64}$Zn isotope in a thermal neutron flux of a nuclear reactor. Table 1 shows the thermal neutron capture cross sections for different zinc isotopes.

Table 1. Isotopic composition and thermal neutron capture cross sections

| Isotope of zinc | 64 | 65 | 66 | 67 | 68 | 70 |
| --- | --- | --- | --- | --- | --- | --- |
| Content in natural Zn (%) | 48.6 | 0 | 27.9 | 4.1 | 18.8 | 0.62 |
| Content in enriched Zn (%) | 94 | 0 | 6 | 0 | 0 | 0 |
| Thermal neutron capture cross section (b) | 0.787 | 64.03 | 0.618 | 7.47 | 1.065 | 0.0917 |

$^{64}$Zn isotope has a low neutron capture cross section, so for the production of a large amount of $^{65}$Zn is required irradiate a large mass of $^{64}$Zn. In turn, the resulting $^{65}$Zn has a high neutron capture cross section and it will be significantly "burning out" in the reactor. Accumulation of $^{65}$Zn in reactors is described by the expression:

$$N_{65}(t) = \frac{N_{64}(0) \cdot \Phi \cdot \sigma_{64}}{\Phi \cdot (\sigma_{64} - \sigma_{65}) - \lambda} \cdot (e^{-(\lambda + \Phi \sigma_{65})t} - e^{-\Phi \sigma_{64} t}) \quad (4).$$

Here $N_{64}(0)$ is the initial number of the $^{64}Zn$ atoms in the neutron flux $\Phi$; $\sigma_{64}$ and $\sigma_{65}$ – neutron capture cross sections on the $^{64}Zn$ and $^{65}Zn$ isotopes; $\lambda$ – $^{65}Zn$ decay constant.
Figure 5 shows the accumulation curve $N_{65}(t)$.

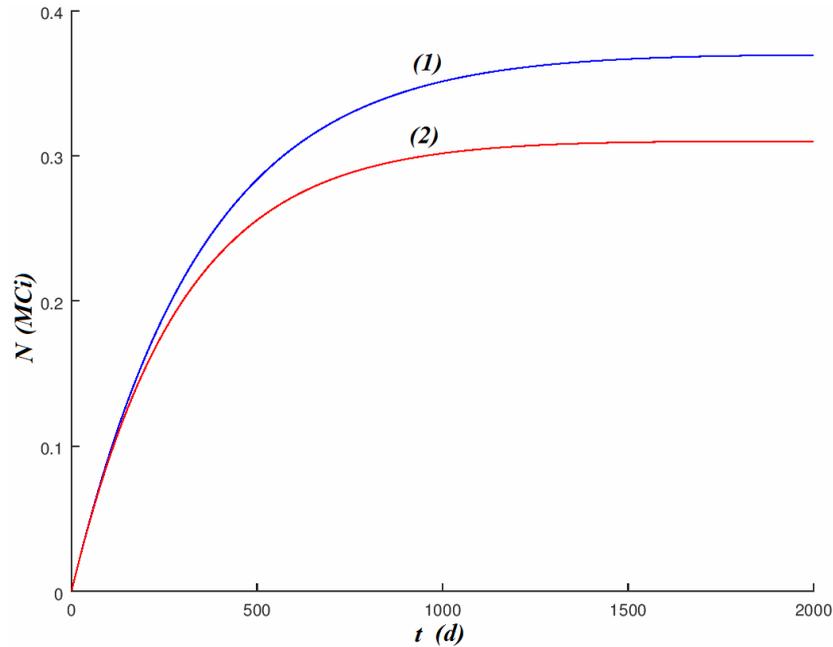

Fig.5. The curves of the accumulation amount of the 65Zn activity $N_{65}$ in the thermal neutron flux $\Phi = 1 \cdot 10^{14}$ cm$^{-2}$sec$^{-1}$ for 20 kg of 94 %- enriched $^{64}Zn$. (1) – without "burnout" of $^{65}Zn$; (2) – with "burnout" of $^{65}Zn$.

A maximal activity of about 0.3 MCi can be attained with 20 kg of 94%-enriched $^{64}Zn$ in a reactor with a thermal neutron flux $\Phi = 1 \cdot 10^{14}$ cm$^{-2}$sec$^{-1}$

From the reference [16] the 65Zn burnout cross section (in (n,α) reaction) in the thermal neutron flux is of 250 b; hence, to accumulate needed activity, it is necessary to irradiate zinc mass about 20 % more. All given below estimates are for $\sigma_{65}$=64 b.

The calculations of the $^{65}Zn$ production were performed for the three reactors: MIR (JSC «SSC SRIAR», Dimitrovgrad, Ulianovskaja obl.), IVV-2M (JSC «IRM», Zarechnyj, Sverdlovskaja obl.) and L-2 (FSUP «PU «Majak», Ozersk, Cheljabinskaja obl.).

For the MIR research reactor, the possibilities of the $^{65}Zn$ accumulation in irradiation of two types of target material – zinc oxide (ZnO) and zinc metal (Zn) enriched with 64 isotope – with various placement of target in the irradiation device were studied.

Activity 0.33 MCi $^{65}Zn$ can be reached by irradiating of 32 kg (6.4 liters) ZnO oxide in 295 days; by increasing the mass of the oxide to 42 kg (8.4 liters), the duration of irradiation is reduced by half, to 150 days. The same activity (0.33 MCi) on metallic zinc can be obtained at a zinc mass of 33 kg (4.7 liters) irradiation during 230 days or 40 kg (5.7 liters) during 170 days.

The activity of 0.5 MCi can be reached with the same mass of irradiated material. To do this, it is necessary to increase the duration of irradiation up to 2 years for 32 kg of ZnO and for 33 kg of metallic zinc, or to increase up to 480 days for 42 kg of ZnO or 40 kg of metallic zinc.

Calculations of $^{65}Zn$ production in the IVV-2M reactor showed the possibility of obtaining 0.5 MCi activity by irradiating of 30 kg metallic zinc for 460 days or 40 kg of zinc oxide enriched up to 94% by $^{64}$Zn - during 490 days. For 0.33 MCi activity will need approximately 1.5 times less of irradiation materials.

According to the calculations of the $^{65}$Zn production in the reactor L-2 in the Mayak: by irradiation of zinc with 94 % enrichment in the isotope $^{64}$Zn over 200 days is possible to reach a specific activity of 9.5 Ci/g Zn. The calculations assume the use zinc oxide tablets ZnO with density of about 5 g/cm$^3$ (zinc oxide density is of 4 g/cm$^3$). In this case, to obtain the activity of 0.33 MCi, it is necessary to irradiate in the reactor about 34.7 kg of metallic zinc or 43 kg of ZnO. The volume of the active part of the source in this case is 8.7 liters, i.e. ~ 15 times more than the volume of the $^{51}$Cr source in the BEST experiment.

Thus, the active part of the 0.33 MCi $^{65}$Zn source will have dimensions from 3.0 to 8.7 liters, depending on reactor. Let us consider how the source dimensions affect the quality of the BEST-2 measurements, comparing the $^{65}$Zn source measurements with dimensions of 8 liters with the $^{51}$Cr source measurements with volume of 0.6 liter.

## 6. Source size

In the BEST-2 experiment the source will be placed in the center of both target zones along a tube with an inner diameter of 21 cm. Therefore, the diameter of a cylindrical source should not exceed 19 cm. To ensure the safety of personnel the source will be shielded by 1.5 cm thick tungsten. Then the diameter of the active part of the source will be about 16 cm, and its height is about 40 cm for activity of 0.3 MCi and volume of 8 liters. Fig. 6 shows the functions $s(L)$ - the relative neutrino registration probability in the target at distance $L$ from the point of birth in the source, which bottom is shifted 10 cm below the center of the target. Compared to the functions $s(L)$ obtained for the chromium source size, for a zinc source with larger dimensions there is a large area of intersection of lengths $L$ for both target zones.

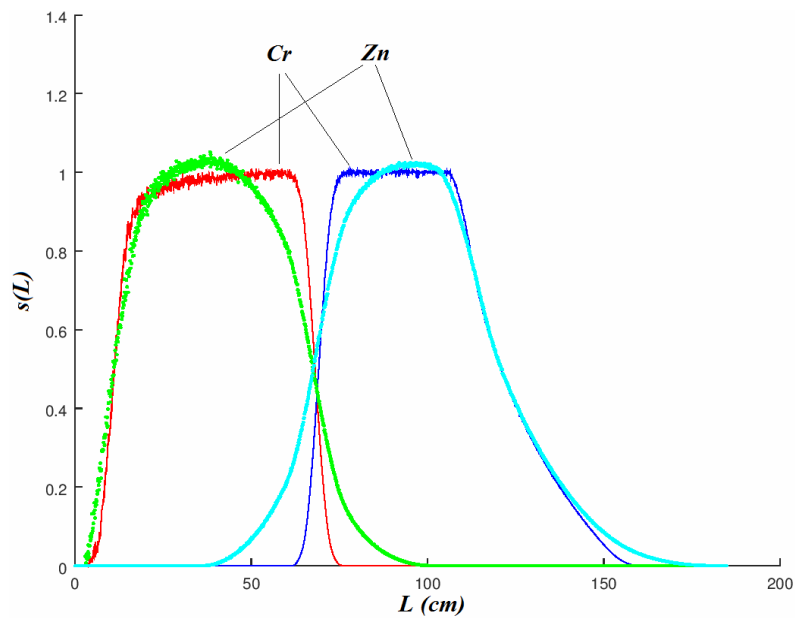

Fig.6. Dependencies of the relative probabilities of neutrino registration $s(L)$ in the inner (spherical) and outer (cylindrical) zones of the gallium targets on the path length $L$ from the point of birth for the cylindrical sources $^{51}$Cr (ø8.6 × 9.5 cm) and $^{65}$Zn (ø16 × 40 cm).

Figure 7 shows the ratio of the expected capture rate in the two zones of target ($R_{in}/R_{ex}$) for the $^{51}$Cr and $^{65}$Zn sources, depending on the value of the parameter oscillations $\Delta m^2$ for the case $\sin^2 2\theta = 0.30$. The regions of sensitivity of $\Delta m^2$ determination for two sources are shifted relative to each other so that the maxima and minima of one curve fall on the values equal to one, of the second curve, that means different sensitivity of experiments with these sources to the possibility of $\Delta m^2$ parameter determination. The amplitude of the first extremes of the $^{65}$Zn curve is about 15% lower than for $^{51}$Cr, and this difference in sensitivity can be reduced by decreasing the size of the source.

In one experiment, BEST or BEST-2, the most probable $\Delta m^2$ values determined from the experiments correspond to the extremes of the curves shown in Fig.7. In the case of two experiments, the most probable value $\Delta m^2$ will be determined according to the weight in each experiment and can take any value in the corresponding range of values.

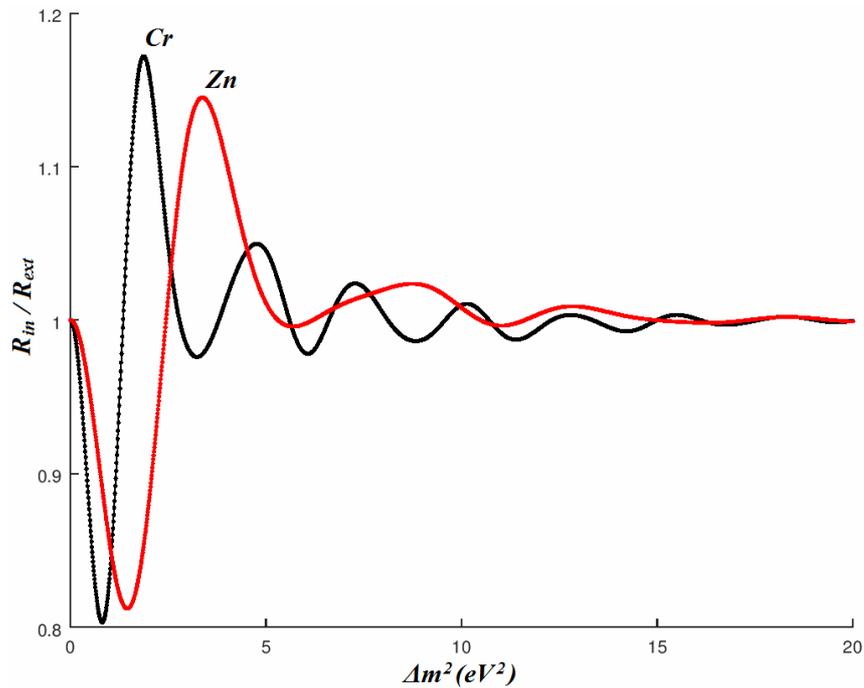

Fig.7. Dependency of ratio of the expected neutrino capture rates in the two zones of target ($R_{in}/R_{ex}$) from the oscillation parameter $\Delta m^2$ for the case $\sin^2 2\theta = 0.30$ for the $^{51}$Cr and $^{65}$Zn sources.

## 7. Conclusion

In the article the questions of performing the experiment BEST-2 to search for oscillations of electron neutrinos from the source of $^{65}$Zn in the sterile state on the two-zone gallium target is considered. The experiment BEST-2 is a natural continuation of the experiment BEST with the 3 MCi $^{51}$Cr source. Experiment will be carried out on the same equipment according to the same scheme. For comparable sensitivity to oscillations, the $^{65}$Zn source can have an activity of 0.33 MCi. The 65Zn source can be made by irradiation with thermal neutrons in MIR-type reactors (JSC "SSC NIIAR", Dimitrovgrad, Ulyanovsk region), IVV-2M (JSC "IRM", Zarechny, Sverdlovsk region) or L-2 (FSUE "PO "Mayak", Ozersk, Chelyabinsk region) for 200 to 490 days. The source dimensions (from 3.0 to 8.7 liters) allow to carry out qualitative measurements to search for oscillations on a short basis on a two-zone gallium target of the BEST experiment.

The BEST and BEST-2 experiments with comparable sensitivity to oscillations supplement each other in determining the parameter of oscillations $\Delta m^2$: due to the difference of the neutrinos energies from the two 51Cr and 65Zn sources the sensitivity zones for determining the parameter $\Delta m2$ in the two experiments are shifted so that the maximum of one experiment sensitivity approximately falls on the minimum sensitivity of the second one and vice versa. The combined result of both experiments will give not only the best experimental statistics of measured oscillations (if they are in the search area), but also - the possibility to measure precisely the parameters of these oscillations.

The work was performed using the scientific equipment of UNU GGNT BNO INR RAS with partial financial support of the Ministry of education and science of the Russian Federation: agreement № 14.619.21.0009, unique identifier of the project RFMEFI61917X0009.
The work by the staff of the Institute for Nuclear Research of the Russian Academy of Sciences related to the estimation of the required activity of the source $^{65}$Zn, its dimensions and production capabilities, performed with the financial support of RFBR within the framework of the scientific project № 17-02-00690.